\documentclass[graybox]{svmult}

\usepackage{mathptmx}       
\usepackage{helvet}         
\usepackage{courier}        
\usepackage{type1cm}        
\usepackage{makeidx}         
\usepackage{graphicx}        
\usepackage{multicol}        
\usepackage[bottom]{footmisc}
\usepackage{wasysym}

\usepackage{epsf}

\makeindex             

\begin{document}

\title*{Quantum dimer models and exotic orders}
\author{K. S. Raman, E. Fradkin, R. Moessner, S. Papanikolaou and S. L. Sondhi}
\institute{K. S. Raman \at Department of Physics and Astronomy, University of California at Riverside, Riverside, CA 92521, \email{kumar.raman@ucr.edu}
\and E. Fradkin \at Department of Physics, University of Illinois at Urbana-Champaign, 1110 West Green Street, Urbana, IL 61801, \email{efradkin@uiuc.edu}
\and R. Moessner  \at Rudolf Peierls Centre for Theoretical Physics, Oxford University, Oxford, OX1 3NP, UK, \email{moessner@thphys.ox.ac.uk}
\and S. Papanikolaou \at Department of Physics, University of Illinois at Urbana-Champaign, 1110 West Green Street, Urbana, IL 61801, \email{papanikl@uiuc.edu}
\and S. L. Sondhi \at PCTP and Department of Physics, Princeton University, Princeton, NJ 08544 \email{sondhi@princeton.edu}}

%
%
\maketitle

\abstract*{We discuss how quantum dimer models may be used to provide ``proofs of principle" for the existence of exotic magnetic phases in quantum spin systems.  The material presented here is an overview of some of the results of Refs.~\cite{rms05} and \cite{prf06}.} 

\abstract{We discuss how quantum dimer models may be used to provide ``proofs of principle" for the existence of exotic magnetic phases in quantum spin systems.  The material presented here is an overview of some of the results of Refs.~\cite{rms05} and \cite{prf06}.}

\section{Introduction}
\label{sec:1}

Consider a system of quantum spins on a lattice with antiferromagnetic interactions.  The energy of a given pair of spins is minimized by forming a singlet but as a spin can form a singlet with only one other spin, the system is unable to simultaneously optimize all of its interactions.  In this sense, the system is said to be {\em frustrated}.  The most common solution to this problem is for the spins not to form singlets but instead to magnetically order, as in the classical Neel state.  However, one may also envision non-magnetic states where the order is contained in how the singlet pairs organize.  In this paper, we concentrate on phases where the singlets form between nearest-neighbor spins.  Such phases are relevant in systems having a spin gap.      

A nearest-neighbor singlet, also called a short-range valence bond, may be represented by drawing a dimer across the corresponding link of the lattice.  Arrangements of singlet pairs correspond to dimer coverings of the lattice, where the hard-core condition of one dimer per site captures the frustration.  This picture may be abstracted to a low energy description where the fundamental degrees of freedom are the dimers themselves.  Quantum dimer models\cite{sr-rvb87, Rokhsar88} describe how exotic phase diagrams 
can arise from the competition between quantum fluctuations of the dimers and various potential energies and are, perhaps, the simplest models containing the physics of frustration.

The suggestion that microscopic frustration may lead to nontrivial phases such as incommensurate crystals\cite{bak82} or liquids of resonating singlets\cite{fazekasanderson} is a rather old one.  The topic is of current interest due to proposals that such phases may be relevant to understanding high $T_c$ superconductors and other strongly correlated systems\cite{andersonrvb, AtoZ04, stripes}.  In this context, we may ask how a particular exotic phase can, in principle, arise from a microscopic Hamiltonian that is local and does not break any lattice symmetries.  One way to answer such questions of principle is to construct an explicit model having the phase in question.  Quantum dimer models are well suited for this purpose because of the relative simplicity of the dimer Hilbert space.  This leads to the question of how these effective models can arise from models with more physical degrees of freedom, such as SU(2) invariant spin models.   

In section \ref{sec:qdm}, we review some facts about the simplest dimer models.  In section \ref{sec:stairs}, we outline the construction of a dimer model that contains a devil's staircase of crystalline phases, including states with arbitrarily long (and hence incommensurate) periods.  In section \ref{sec:su2}, we discuss how dimer models may be mapped onto SU(2) invariant spin models on decorated lattices.  In particular, the existence of a liquid phase of resonating dimers in the simplest triangular lattice dimer model implies the existence of a liquid phase of resonating singlets on a decorated triangular lattice.  This paper gives an overview of some of the results in Refs.~\cite{rms05} and \cite{prf06}.  

\section{Basic models}
\label{sec:qdm}
\begin{figure}[ht]
{\begin{center}
\includegraphics[width=3in]{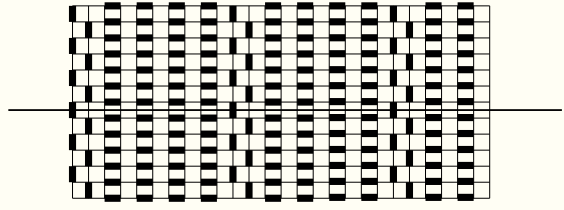}
\caption{Winding numbers for the square lattice case.  The thick horizontal line extends around the lattice.  If the vertical lines of lattice are labeled alternately A and B, then the winding number is 
defined as $N_A-N_B$ where $N_{A,B}$ is the number of dimers crossing the thick line along A,B lines.  For the section of the lattice shown, $N_A=3$ and $N_B=0$.  One may verify that local dimer rearrangements will not change this number.  In the triangular lattice case, the relevant quantity is defined by just counting the number of dimers intersecting this reference line.  Whether this number is odd or even is a topological invariant but the value of the number can change with local rearrangements of the dimers.}
\label{fig:tilt}
\end{center}}
\end{figure}

The Hilbert space of quantum dimer models is defined by taking each dimer covering of the lattice as a basis vector.  The inner product is usually defined by taking different dimer coverings to be orthogonal.  The dimer Hilbert space may be partitioned into topological sectors labeled by a pair of winding numbers (Figure \ref{fig:tilt}).  The winding numbers are global invariants in that they are not affected by local rearrangements of dimers.  In particular, local operators (including the Hamiltonian) will not have matrix elements between states in different sectors.  For bipartite lattices, the number of sectors is extensive while for non-bipartite lattices, there are four sectors corresponding to each winding number 
being either even or odd.  

The original quantum dimer model of Rokhsar and Kivelson\cite{Rokhsar88} was defined on the square lattice with the Hamiltonian:
\begin{equation}
\scalebox{1.0}{\includegraphics[width=3in]{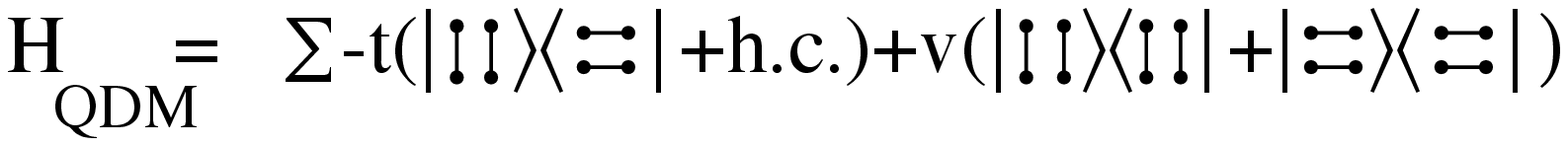}}\\
\label{eq:hqdm}
\end{equation}
The two terms are projection operators that flip (the $t$ term) or count (the $v$ term) the dimers if the plaquette in question has parallel dimers, which we call a ``flippable" plaquette, and annihilate the state otherwise.  The sum is over all plaquettes in the lattice.  The model may be generalized to other lattices where the flippable plaquettes now correspond to the dimers occupying alternate bonds of the minimal (even-length) resonance loop in the lattice.  For 
the triangular lattice\cite{MStrirvb}, these loops are still length 4 but there are now six distinct ways of having parallel dimers instead of just two for the square lattice.  For the honeycomb lattice, the loops are length six and the projection operators are three-dimer moves analogous to the benzene resonance.  For the pentagonal lattice\cite{rms05}, the loops are length eight, the operators are four-dimer moves and so on.

Models of the Rokhsar-Kivelson type have certain generic properties.  When $v>t$, 
Eq.~\ref{eq:hqdm} is positive definite so the system seeks to minimize the number of flippable plaquettes.  In particular, many lattices, including the ones mentioned above, may be covered without having any flippable plaquettes and such ``staggered" states will be zero energy ground states in this limit.  When $-v\gg t$, the system seeks to maximize the number of flippable plaquettes so selects an analog of the ``columnar" state.  When $v=t$, Eq.~\ref{eq:hqdm} is again positive definite, but in addition to the staggered states, there are a number of liquid states which also have zero energy\cite{Rokhsar88}.  The liquid wavefunctions may be written as equal amplitude superpositions $|\psi\rangle=\sum_c |c\rangle$  where the sum is over all states that may be connected by repeatedly applying the flip (kinetic) term in Eq.~\ref{eq:hqdm}.  There will be at least one such state in every topological sector.  For the square lattice, there is only one state in each sector because the flip term is believed to be ergodic within the sectors but additional subtleties may appear in other lattices\cite{MStrirvb}.  Because these are zero energy states, any combination of them will be equally valid as a ground state, including the equal amplitude superposition of {\em all} dimer coverings.

The liquid states at $v=t$, which is called the RK point, have no local order but they do carry the global quantum numbers of their topological sectors.  However, the nature of the liquid states depends intimately on the lattice geometry.  For some bipartite lattices, including the square and honeycomb, these states have algebraically decaying correlations, are gapless, and the RK point is a critical point separating crystalline phases with minimal and maximal winding number.  Field theoretic studies\cite{vbs04,fhmos04} suggest that perturbations about this critical point can stabilize crystalline phases with intermediate winding number and, generically, a devil's staircase of such states will occur in between the states of minimal and maximal winding.  In particular, it was noted that incommensurate crystals\cite{incomm} could be obtained by suitable tuning of the perturbation.   

For some non-bipartite lattices, including the triangular and kagome, the correlations decay exponentially, the liquid states are gapped, and the RK point is the edge of a liquid 
phase\cite{MStrirvb}.  This was the first example of a model containing a liquid phase of resonating bonds and validated earlier suggestions of a physical mechanism by which such a phase could be stabilized, namely geometric frustration enhanced by strong ring exchange\cite{gregMSE}.  For the kagome lattice, a simpler dimer liquid phase, with exactly zero correlation beyond one lattice spacing, was obtained in Ref. \cite{gregMSP} using a different construction with only flip terms.  

In the next section, we discuss how a different kind of exotic structure can be realized in a dimer model.  Then, in section \ref{sec:su2}, we show how these proofs of principle in dimer models may be extended to SU(2)-invariant spin systems. 

\section{Modulated states and the devil's staircase}
\label{sec:stairs}

We now outline a dimer model that shows a devil's staircase of crystalline phases where the period becomes arbitrarily large\cite{prf06}.  The following construction is valid in the limit of strong coupling and weak (quantum) fluctuations, in contrast to the RK point which corresponds to the opposite limit of strong fluctuations.  In this sense, the connection between the following devil's staircase and the one predicted by field theory to occur near bipartite RK points (mentioned above) is tempting but speculative.   
Similarly, we work on the square lattice for convenience but we do not believe this choice is so important in the strong coupling regime.

Our strategy is to perturb a model that contains a degenerate point separating crystalline states of different winding number.  In particular, we consider the following diagonal Hamiltonian:
\begin{equation}
\scalebox{1.0}{\includegraphics[width=4.5in]{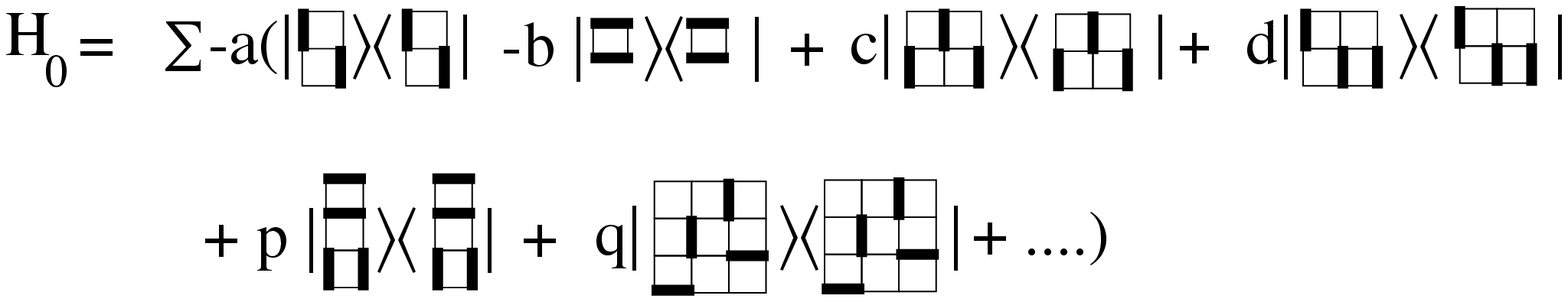}}\\
\label{eq:parent}
\end{equation}
where the dots denote terms related by lattice symmetries to ones shown, which will appear with the same coefficient.  Note that Eq.~\ref{eq:parent} is local and does not break lattice symmetries.   The coefficients satisfy $p,q>c,d>a,b>0$ but fine tuning is not required.  Terms $a$ and $b$ are competing attractive interactions while the remaining terms are repulsive.  This Hamiltonian is designed to favor states where every dimer participates in only two attractive bonds of the same type.  Terms $c$ and $d$ penalize arrangements with dimers participating in more than two attractive bonds\cite{stagg}.  Terms $p$ and $q$ serve a technical purpose discussed in Ref.~\cite{prf06}.   

\begin{figure}[ht]
{\begin{center}
\includegraphics[width=3in]{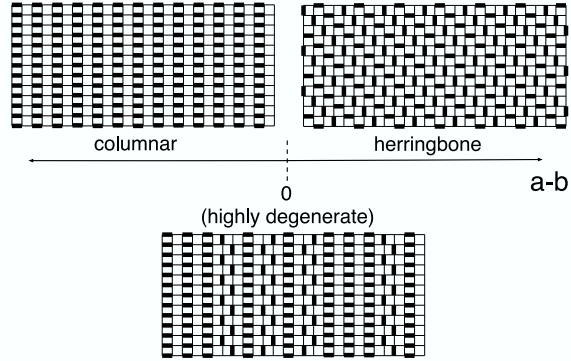}
\caption{Ground state phase diagram of the parent Hamiltonian $H_{0}$ as a 
function of the parameter $a-b$ (see Eq.~\ref{eq:parent}).  When $a-b=0$, there are 
a number of degenerate ground states.  The maximally staggered 
configuration is commonly called the ``herringbone'' state.}
\label{fig:phase0}
\end{center}}
\end{figure}

The quantum system described by the Hamiltonian in Eq.~\ref{eq:parent} has the zero temperature phase diagram shown in Fig.~\ref{fig:phase0}.  As the figure shows, the system prefers exclusively $a$ or $b$ bonds depending on which coefficient is larger.   When $a=b$, there is a large degeneracy where the preferred states involve thin staggered domains separating columnar regions of opposite orientation; the staggered domains come in two orientations as illustrated in the figure.

We perturb this model with a resonance term that is equivalent to two actions of the flip term in Eq.~\ref{eq:hqdm}:
\begin{equation}
\scalebox{1.0}{\includegraphics[width=3.8in]{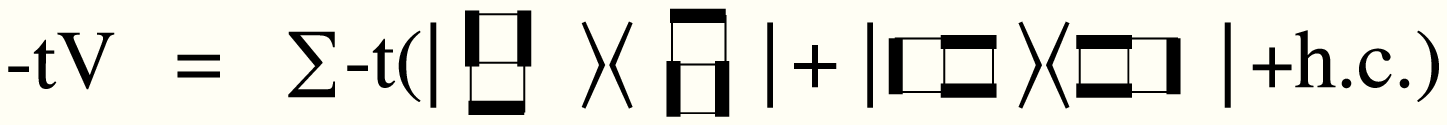}}\\
\label{eq:V}
\end{equation}
where $0<t \ll a,b$.  We now describe how the existence of the devil's staircase follows by perturbing in $t$ at successively higher orders.  Before doing this, we remark that the main reason for choosing a somewhat complicated form for Eq.~\ref{eq:parent} and a non-standard resonance term (i.e. Eq.~\ref{eq:V}) is that these choices simplify the calculational details.  As discussed in Ref.~\cite{prf06}, we anticipate that this construction can be made to work for simpler diagonal Hamiltonians and with the familiar two-dimer resonance. 

Referring to Fig.~\ref{fig:phase0}, we see that the operator \ref{eq:V} does not affect the phases on either side of the degenerate point but will partially lift the degeneracy at $a=b$.  In particular, this term causes the staggered domain walls to fluctuate which stabilizes them at second order in perturbation theory. 
In this sense, our model is a quantum analog of the Pokrovsky-Talapov model of fluctuating domain walls\cite{poktal78}.  In that classical model, thermal fluctuations give entropy to the domain walls and the competition between this and their energy cost induces a commensurate-incommensurate transition\cite{bak82} to a striped phase where the walls arrange periodically with spacing that can be large compared to the interaction scale.  In the present case, quantum fluctuations play the role of temperature and cause modulated phases to appear. 

\begin{figure}
    \begin{tabular}{ccc}
	\centering
	\begin{minipage}{1.5in}
	    \includegraphics[width=1.5in]{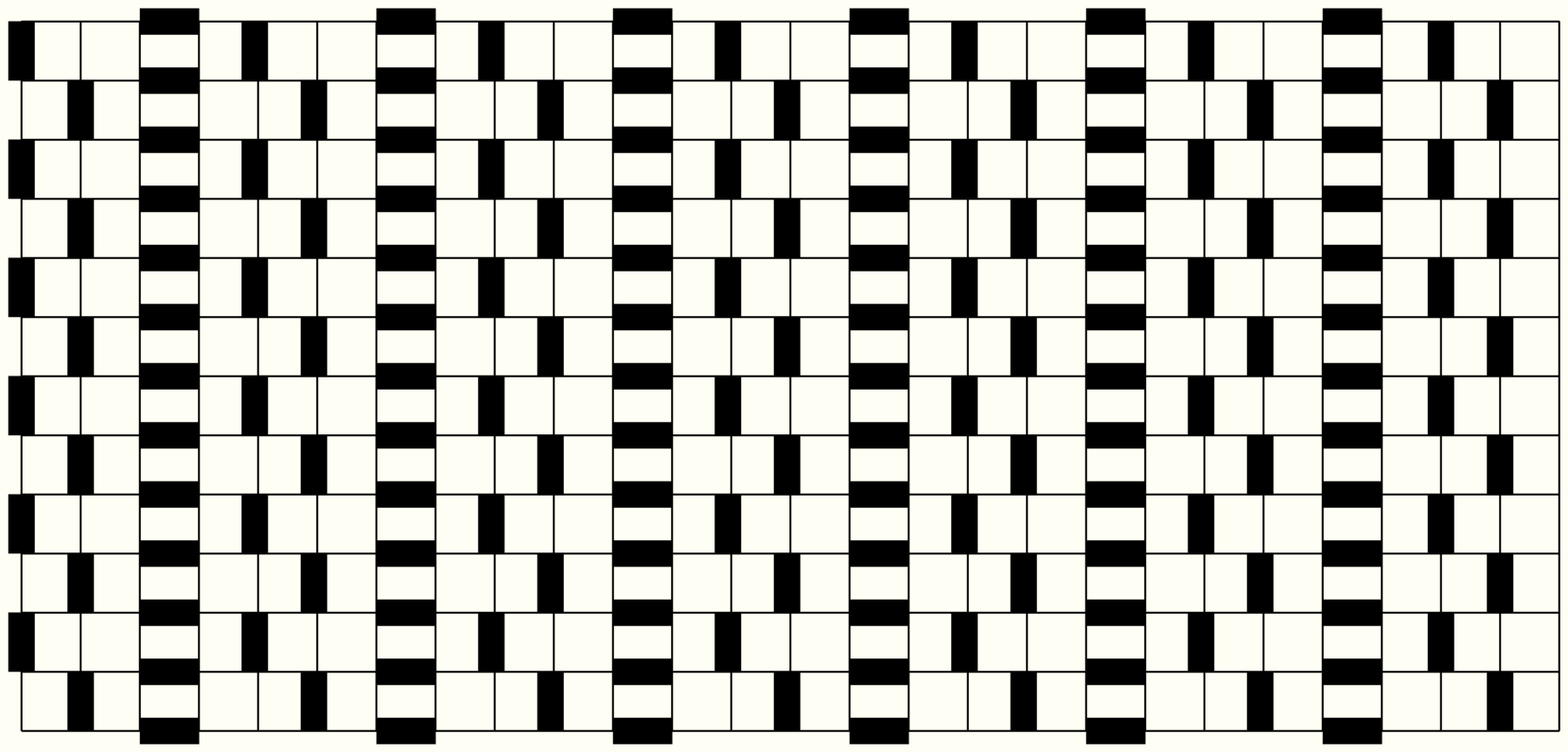}
	 \end{minipage}&
	 \begin{minipage}{1.5in}
	    \includegraphics[width=1.5in]{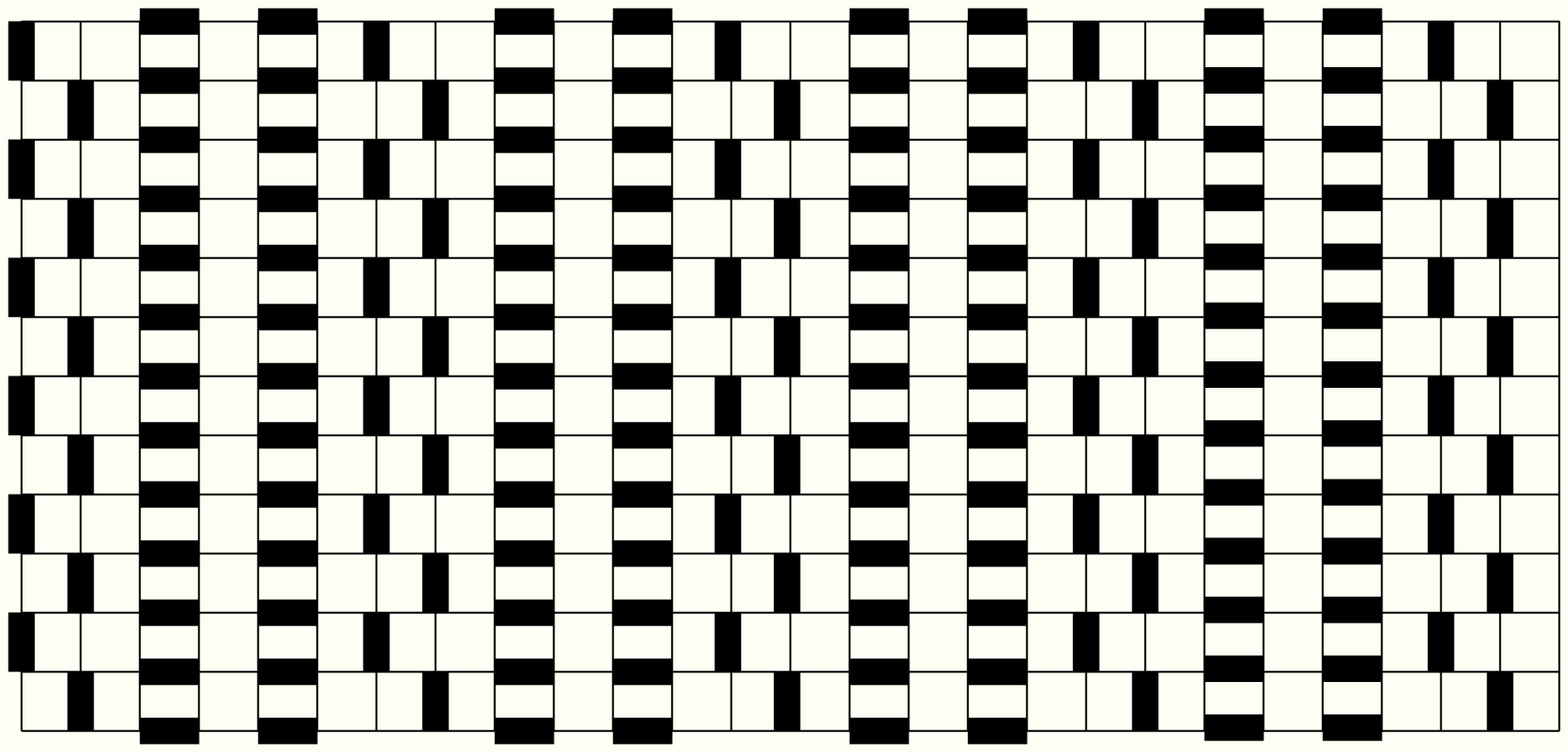}
	 \end{minipage}&
	 \begin{minipage}{1.5in}
	   \includegraphics[width=1.5in]{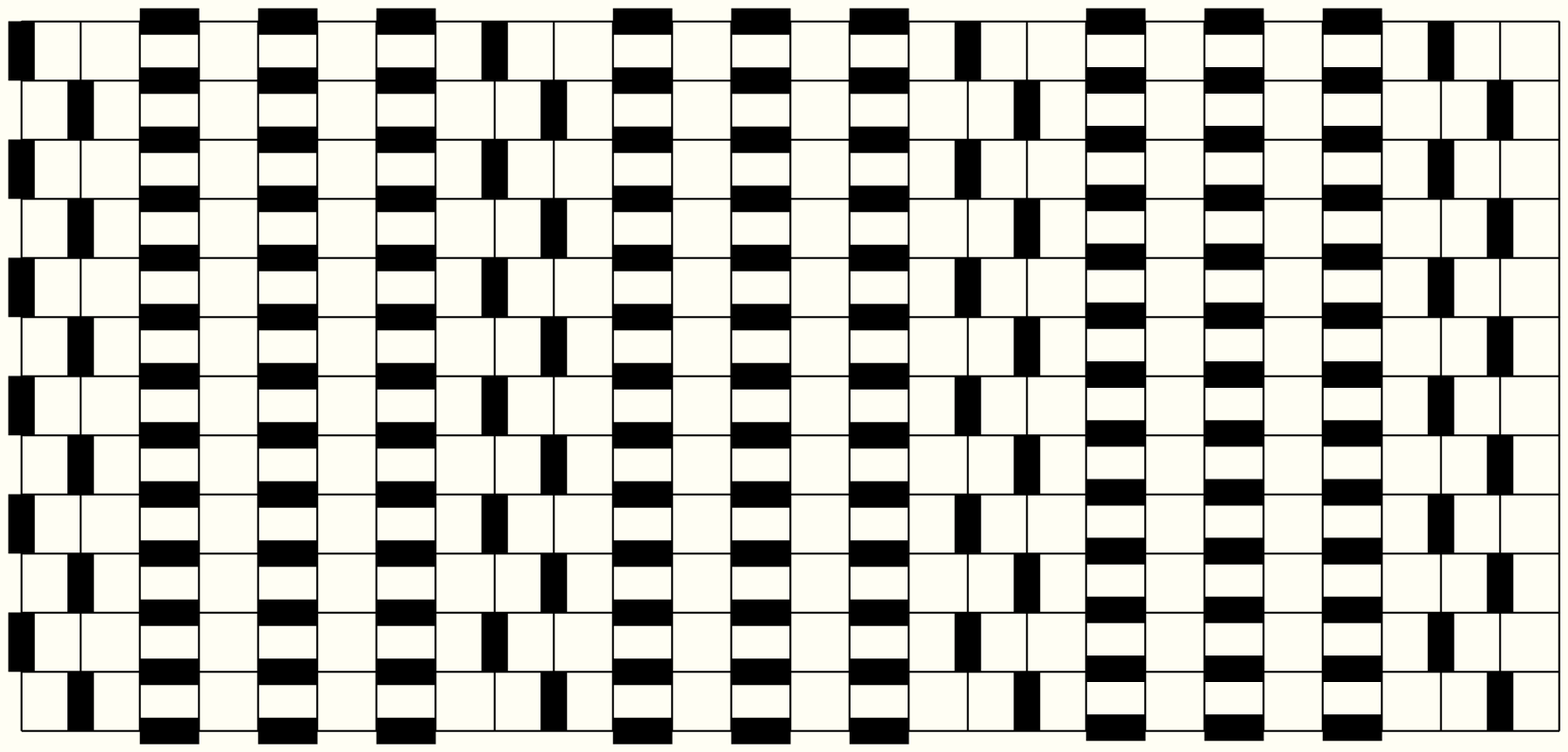}
	  \end{minipage}\\ (a) & (b) & (c) \\	 
    \end{tabular}	 
\caption{The [11], [12], and [13] states involve staggered domains separated by one, two, and three columnar units respectively.  In these states, the staggered domains have the same orientation.  
We can similarly define [14], [15] etc.}
\label{fig:ideal}
\end{figure}

A technical issue is that the extent to which a staggered domain is stabilized by second order processes depends on its environment.  One consequence of this is the favoring of states, such as the [1n] sequence (Figure \ref{fig:ideal}), where the staggered domains have the same orientation.  Also, while the [11] state clearly has the maximum number of resonances, these resonances are individually weaker than those in the [12] state because of the higher energy virtual states that are involved (due to term $c$ in Eq.~\ref{eq:parent}).  If the [11] phase is selected when $a=b$, then by slightly detuning from this point (by an amount of order $t^2$), we will reach a value of $b$ where the ``self-energy" of a columnar line \cite{selfenergy} becomes degenerate with the energy of a resonance-stabilized staggered  line in the [11] state.  However, $b$ must be further increased before the advantage of [12] staggered lines is similarly compensated.  Therefore, both the [11] and [12] phases appear in regions of width $~\sim t^2$ between the initial phases and to second order in $t$, we obtain the phase diagram in Fig. ~\ref{fig:phase}a.  Note that the winding number increases as we move from left to right in the phase diagram.

\begin{figure}[ht]
{\begin{center}
\includegraphics[width=3in]{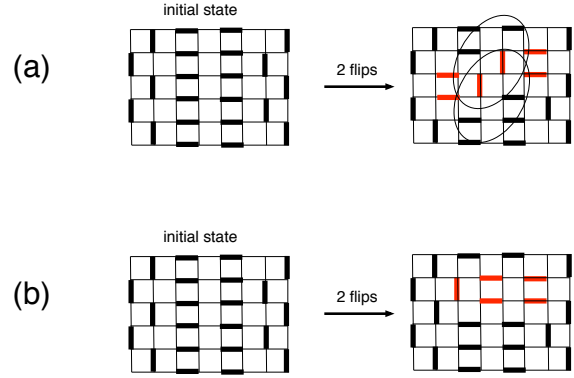}
\caption{The fourth order resonances driving the transition.  We emphasize that these figures represent terms that appear in fourth order perturbation theory, {\em not} additional terms added to the Hamiltonian.  The circled cluster in (a) refers to term $q$ in Eq.~\ref{eq:parent}.}
\label{fig:4res}
\end{center}}
\end{figure}

\begin{figure}[ht]
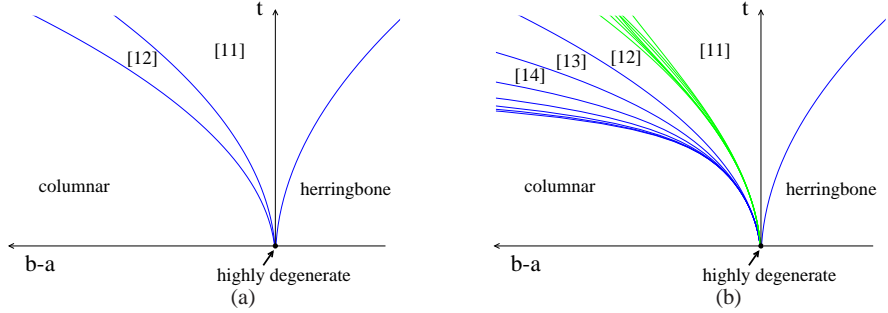

    \begin{tabular}{cc}
	\centering
	\begin{minipage}{2.5in}
	    \includegraphics[width=2.1in]{phase2.eps}
	 \end{minipage}&
	 \begin{minipage}{2.5in}
	    \includegraphics[width=2.1in]{phasefull.eps}
	 \end{minipage}\\ (a) & (b)\\	 
    \end{tabular}	 
    \caption{(a) Ground state phase diagram of $H=H_{0}+tV$ from second order
perturbation theory.  The width of the [11] and [12] phases are order $t^{2}$. 
(b) Sketch of the ground state phase diagram to all orders in perturbation theory.  While the opening of the [11]-[12] boundary is explicitly indicated, the other boundaries will also open.  The collection of phases forms a devil's staircase.  Note:  These figures are not the result of a simulation but a sketch of the general properties of the phase diagram.}
\label{fig:phase}
\end{figure}

On the [11]-[12] and [12]-columnar boundaries, the phases and states with intermediate winding are degenerate to order $t^2$ but these degeneracies will be lifted at higher orders.  Considering the [12]-columnar boundary at fourth order in perturbation theory, the relevant resonances are shown in Figure \ref{fig:4res}.  Process (a) destabilizes the [12] state due to the high energy virtual state involving term $q$ in Eq.~\ref{eq:parent} (there is an analogous process involving term $p$) and its contribution is larger than process (b) which is stabilizing. In the [13] state, the analog of process (a) will not contribute because the flipped clusters are disconnected which means that the energy and wavefunction terms in the perturbation series exactly cancel.  However, process (b) will occur and stabilize a [13] phase in a region of width $\sim t^4$ between the [12] and columnar states.  The argument may be applied inductively.  Along the 
[1n]-columnar boundary, at (2n)th order in the perturbation, the analog of the Figure \ref{fig:4res}a will destabilize the [1n] state relative to states of lesser winding while the analog of Figure \ref{fig:4res}b selects the [1,n+1] state from this set.  The new [1,n+1] phase will occupy a region of width $\sim t^{2n}$ in the phase diagram and so on.  While higher orders in the perturbation theory involve increasingly complicated resonances, including fluctuations causing the staggered lines to effectively ``break apart", the calculation is designed so that these terms amount to self-energy corrections that simply move the boundaries.  The stabilization of phases in the [1n] sequence will always be governed by the analogs of the straight resonances in Figure \ref{fig:4res}.        

The [11]-[12] boundary can similarly open at higher orders and in Ref.~\cite{prf06}, we verify that a [11-12] phase (the notation means that the repeating unit is a staggered domain followed by one columnar unit followed by another staggered domain followed by two columnar units) is stabilized at sixth order between the [11] and [12] states.  The finer boundaries can also open and we have also verified that a $[11-(12)^2]$ phase occurs at eighth order between the [11-12] and [12] phases.  We are less certain about this fine structure because the resonances involved are more complicated than in the primary [1n] sequence.  However, we may speculate that the opening of boundaries will continue to finer scales, at least for some range of parameters.  In this sense, the phase diagram will most generically be described by an incomplete devil's staircase, as shown in Figure \ref{fig:phase}b.  Moving from left to right in Figure 5, the system passes through a series of crystalline states with increasing winding number via first-order phase transitions.   We note that this phase diagram and the structure of the calculation is similar to the 3D classical ANNNI model\cite{fishselke80, bakvonboehm80, yeomans95}.

\section{SU(2)-invariant realizations} 
\label{sec:su2}

Having discussed some examples of how exotic phases can arise in dimer models, we now turn to the question of what this implies for systems with more realistic degrees of freedom.  The interpretation of a dimer as a nearest-neighbor singlet or valence bond involves two simplifications.  The dimers have no orientation and different dimer coverings are orthogonal by definition.  In contrast, valence bonds are oriented\cite{orient} and different valence bond coverings are not orthogonal.  There is also the issue of projecting the spin Hilbert space onto the much smaller dimer Hilbert space.

In this section, we discuss one way in which these issues may be resolved allowing the dimer model phase diagrams to be realized in SU(2)-invariant systems.  We concentrate on the square lattice but the arguments may be generalized to other lattices and higher dimensions.  The interested reader may consult Ref.~\cite{rms05} for details.  We would also like to point out a complementary approach by Fujimoto\cite{fujimoto05}.    

\begin{figure}[ht]
{\begin{center}
\sidecaption
\includegraphics[width=2.2in]{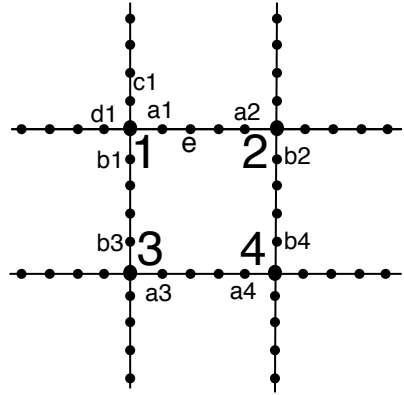} 
\caption{Decorated square lattice for the case $N=4$.}
\label{fig:squaredec}
\end{center}}
\end{figure}

We construct the spin model on a modified square lattice where the links are decorated with an even number $N$ of additional sites (Fig.~\ref{fig:squaredec}).  On each lattice site $i$, we define an operator $\hat{h_i}$ that projects the cluster of that spin and its (two or four) neighbors onto its highest spin state\cite{kleinmodel, majghosh, chayeskiv} (either 3/2 or 5/2).  The parent Hamiltonian is a sum of these operators, one for each lattice site:
\begin{equation}
H_0=\sum_i \alpha_i \hat{h_i}
\label{eq:klein}
\end{equation}
where $\alpha_i$ is a positive constant that, in principle, may vary with $i$.  The SU(2)-invariance of Eq.~\ref{eq:klein} may be seen by explicitly writing down the operators.  Referring to the figure,
$\hat{h_{a1}}=S^2-(\frac{1}{2})(\frac{3}{2})$ where $\vec{S}=\vec{S}_{a1}+\vec{S}_1+\vec{S}_e$ and likewise for the other link sites.  Similarly, $\hat{h_1}=[S^2-(\frac{1}{2})(\frac{3}{2})][S^2-(\frac{3}{2})(\frac{5}{2})]$ where $\vec{S}=\vec{S}_1+\vec{S}_{a1}+\vec{S}_{b1}+\vec{S}_{c1}+\vec{S}_{d1}$ and likewise for the other corner sites.  

As these expressions indicate, $\hat{h_i}$ is a product of terms that annihilate the lower spin sectors so only the highest spin state remains.  Among the configurations annihilated by $\hat{h_i}$ is the case
where two spins in cluster $i$ form a singlet.  This is because the highest spin state must be symmetric
under interchange.  A consequence of all this is that any configuration where every spin is in a singlet with one of its neighbors, which we call a valence bond state, will be annihilated by Eq.~\ref{eq:klein}.  Because $H_0$ is a sum of positive definite operators, its eigenvalues are positive so such annihilated states will be zero energy ground states.  Even a minimal decoration of $N=2$ ensures that the valence bond states (and their superpositions) are the only zero energy states.

In the valence bond states, the chains forming the links of the decorated square lattice are in one of two possible dimerizations.  One of the dimerizations involves the sites of the original square lattice while the other involves only the decorated sites.  In the original square lattice, we may represent the former case by drawing a dimer across the link and the latter by an empty link.  Therefore, the valence bond states correspond exactly with dimer coverings of the lattice but an important technical difference is the non-orthogonality of the valence bond basis.  However, the decoration ensures that the magnitude of the overlap between different valence bond states will always be exponentially small in the decoration $N$.  This allows us to treat the overlap in an expansion that becomes asymptotically exact for large enough $N$.  In contrast to other large-N approaches\cite{largeN}, this procedure occurs entirely within the class of SU(2) models.  

The scale of the spin gap is determined by the minimum of the $\{ \alpha_i \}$.    
If this scale is sufficiently large, then perturbations of Eq.~\ref{eq:klein} will generate effective operators in the valence bond manifold.  In particular, consider the perturbation:
\begin{equation}
    \delta H = J \sum_{\langle ij \rangle} \vec{s}_{i}\cdot\vec{s}_{j} +   v\sum_{\Box}\Bigl(
    (\vec{s}_{1}\cdot\vec{s}_{b_{1}})(\vec{s}_{2}\cdot\vec{s}_{b_{2}})+
    (\vec{s}_{1}\cdot\vec{s}_{a_{1}})(\vec{s}_{3}\cdot\vec{s}_{a_{3}})
    \Bigr)
 \label{eq:pert}
\end{equation}
where the first sum is over nearest neighbors and the second is over square plaquettes (the symmetric terms are not explicitly written).  The degenerate perturbation theory involves accounting for the non-orthogonality of the valence bond manifold.  In particular, if $S_{ij}=\langle i|j\rangle$ is the overlap matrix, then we may consider the orthogonal basis $|\alpha\rangle=\sum_i S^{-1/2}_{\alpha i}|i\rangle$.
Because the overlap between different states is small, we may label $|\alpha\rangle$ by its order unity component.  In terms of this basis, the operator Eq.~\ref{eq:pert} becomes:
\begin{eqnarray}
    H_{\alpha\beta}&=&(S^{-1/2}\delta H S^{-1/2})_{\alpha\beta}\\
    &=&\sum_{ij}(S^{-1/2})_{\alpha i}\langle i|\delta H|j\rangle
    (S^{-1/2})_{j\beta}\nonumber\\&=&    -Jx^{4(N+1)}\Box_{\alpha\beta}+vn_{fl,\alpha}\delta_{\alpha\beta}+O(vx^{4(N+1)}+Jx^{6(N+1)})\nonumber\\ &=&-t\Box_{\alpha\beta}+vn_{fl,\alpha}\delta_{\alpha\beta}+O(vx^{4(N+1)}+tx^{2N})\nonumber\\
\end{eqnarray}
where $x=\frac{1}{\sqrt{2}}$.  $\Box_{ij}$ is a matrix that is 1 if states $|i\rangle$ and $|j\rangle$ differ by the (minimal) length 4 loop and zero otherwise; $n_{fl,i}$ counts the number of flippable plaquettes in state $|i\rangle$; and $t=-Jx^{4(N+1)}$.  

Therefore, up to small corrections, Eq.~\ref{eq:pert} acts like the RK quantum dimer Hamiltonian in the orthogonalized basis and by decorating the lattice with a sufficient number of sites, the matrix elements beyond the dimer model can be made arbitrarily small.  While this procedure will not capture the fine-tuned aspects of the dimer model, such as the $v=t$ critical point for bipartite lattices, without taking $N$ extremely large, the various gapped phases that appear can be obtained for a finite decoration.  In particular, we may repeat this procedure on a decorated triangular lattice to obtain a model with a stable, SU(2)-invariant spin liquid phase.  A similar construction may be used to realize the modulated phases discussed in the previous section.

\section{Conclusion}
\label{sec:concl}

We have shown that incommensurate structures and spin liquids can, in principle, arise from purely 
local Hamiltonians that do not break any symmetries  The usefulness of quantum dimer models in answering this kind of question was hopefully conveyed.  Along with the proofs of principle comes an understanding regarding physical mechanisms by which such structures may form.  In the spin liquid ground states, the key elements were geometric frustration and strong quantum fluctuations through ring exchange.  In the devil's staircase construction, the key ingredients were the fluctuating domain wall picture and competing interactions in a strong coupling limit.  

While the constructions involved rather complicated lattices and/or Hamiltonians, it is likely that the ideas apply for simpler, though perhaps less (analytically) tractable, models where frustration enters in a less rigid way than a hard-core dimer constraint.  For example, we may consider models with longer bonds\cite{moessnersandvik} or doped dimer models\cite{sachdev}.  Therefore, we may hope the ideas presented here will provide a starting point for constructing physically realistic models with exotic phases.

\begin{acknowledgement}
This review is based on a contributed talk by KSR at the PITP-Les Houches School for Quantum Magnetism 2006.  KSR would like to thank the organizers for the opportunity to attend.  This work was supported by the National Science Foundation through the grants NSF DMR0213706 and NSF DMR 0442537.
\end{acknowledgement}

%
%

\begin{thebibliography}{99.}
\bibitem{incomm}
    More precisely, the winding number per unit length (also called the "tilt") of the crystalline state
    could approach a value incommensurate with the lattice.
\bibitem{stagg}
      For example, in the fully staggered state, every dimer would have four attractive $a$ bonds.
\bibitem{orient}
In order to interpet a dimer connecting sites 1 and 2 as a valence bond, we also need to specify the overall sign of the singlet, i.e. $\pm\frac{1}{\sqrt{2}}(1_\uparrow 2_\downarrow - 1_\downarrow 2_\uparrow)$.
\end{thebibliography}
%

\end{document}